\documentclass[12pt]{article}
\usepackage{amsmath,amssymb}
\usepackage{mathbbol,bbm}
\usepackage{epic}

\newcommand{\Cx}{{\mathbb C}}
\newcommand{\Ir}{\mathbb{Z}}

\newcommand{\Rl}{{\mathbb R}}

\newcommand{\idty}{\Eins}

\DeclareMathOperator{\tr}{Tr}
\newcommand{\<}{\langle}
\renewcommand{\>}{\rangle}
\providecommand{\abs}[1]{\lvert#1\rvert}
\providecommand{\norm}[1]{\lVert#1\rVert}

\renewcommand{\c}[1]{\mathcal{#1}}
\newcommand{\g}[1]{\mathfrak{#1}}
\newcommand{\s}[1]{\mathsf{#1}}
\renewcommand{\r}[1]{\mathrm{#1}}

\begin{document}
\setlength{\parskip}{6pt}

\begin{center}
{\LARGE A statistical mechanics view on Kitaev's \\[12pt] 
proposal for quantum memories} \\[12pt]
R.~Alicki$^\dagger$, M.~Fannes$^\ddagger$ and M.~Horodecki$^\dagger$
\\[6pt]
$^\dagger$ Institute of Theoretical Physics and Astrophysics \\
University of Gda\'nsk, Poland \\[6pt]
$^\ddagger$ Instituut voor Theoretische Fysica \\
K.U.Leuven, Belgium
\end{center}

\noindent
\textbf{Abstract:}
We compute rigorously the ground and equilibrium states for Kitaev's
model in 2D, both the finite and infinite version, using an analogy
with the 1D Ising ferromagnet. Next, we investigate the structure of
the reduced dynamics in the presence of thermal baths in the Markovian
regime. Special attention is paid to the dynamics of the topological
freedoms which have been proposed for storing quantum information.

\section{Introduction}

Despite the enormous activity in the field of quantum information in
the last decade, the fundamental problem of how to protect quantum
information for macroscopic periods of time remains open. In contrast
to the popular belief the Fault Tolerant Quantum Computation (FTQC)
ideas~\cite{Shor1996-FT,ZurekFT,DoritFT,Gottesman2000-FT-local,
Preskill1998-rel} do not provide the final solution to this problem.
The mathematical models behind these schemes are highly oversimplified,
phenomenological, and do not take into account the restrictions
imposed by the laws of thermodynamics. As FTQC involves time-dependent
control by external fields the rigorous analysis of the corresponding
models which can be derived from first principles is extremely
difficult. In particular, the standard approximations based on the
rigorous versions of Markovian limits (weak coupling or low 
density~\cite{Davies1974-WCL,Alicki-Lendi,Breuer:book}), which yield 
descriptions consistent with thermodynamics, cannot be applied, 
see~\cite{AHHH2001,AlickiLZ2006-FT,TerhalB-FT,Alicki-TB,AharonovKP-FT-NM} 
in this context. 
Therefore, the approach to quantum
memories based on the idea of self-correcting systems is more
promising for rigorous studies. It deals with $N$-qubit systems with
specially designed Hamiltonians which are supposed to stabilise an
algebra of observables corresponding to one or few qubits, at least
below some critical temperature. The construction of self-correcting
Hamiltonians is based on the properties of error correcting codes and
the most advanced examples are 2, 3 and 4D Kitaev models~\cite{KitaevFT,
DennisKLP2002-4DKitaev} associated with toric codes.

It is instructive to discuss first the mechanism of protection of
classical information using phase transitions in quantum
systems. In the general approach based on the algebraic formalism for
infinite systems, i.e.\ systems in the thermodynamic limit, we
characterise equilibrium states by the KMS~condition~\cite{RB97}. States
that satisfy the KMS~condition are known to have good stability
properties such as passivity~\cite{Pusz-Woronowicz78}. Typically, one expects that
for sufficiently high temperatures there exists a unique KMS~state,
while below a certain critical temperature $T_c$ a simplex of
KMS~states can appear. The extreme states which span this simplex
correspond to pure thermodynamical phases labelled by the
values of the order parameters. They can be used to encode
classical information.

In a complementary dynamical picture we consider an $N$-particle system
weakly interacting with a heat bath at temperature $T$. The finite
system possesses a unique equilibrium state given by a Gibbs density
matrix. However, for systems exhibiting phase transitions and for
$T<T_c$ the unique equilibrium state can be decomposed into many
long-living metastable states corresponding to local minima of the free
energy. These minima are separated by free energy barriers of heights
proportional to $N$. This makes transitions caused by
thermal fluctuations highly improbable and leads to life-times
exponentially increasing with $N$. Obviously, the metastable states
should become equilibrium (KMS) states in the limit $N\to\infty$. 
Although the equivalence of equilibrium and dynamical pictures is
generally accepted there exist only few rigorous results that support
this, e.g.\ the mean-field Curie-Weiss model~\cite{Martin1977-Ising-CW} and
Bose-Einstein condensation~\cite{BuffetSP1984-openbec}.

From the discussion of above it follows that extending these ideas to 
the field of quantum information in order to construct
quantum memories is not possible. The encoding of qubits into different
thermodynamical phases of an infinite system is impossible because of the
simplex structure of the KMS~states. A simplex is a purely classical
state space and cannot even support a single qubit. 

Actually, in the literature one attempts to construct self-correcting
Hamiltonians for $N$ qubits in such a way that their degenerated ground
states correspond to codewords encoding states of 
one or few qubits \cite{KitaevFT,DennisKLP2002-4DKitaev}.
However, good codewords are locally indistinguishable, which implies
that in the thermodynamic limit all of them merge into a single ground
state of the infinite system. Therefore one doesn't expect that such
systems exhibit phase transitions at any temperature and so cannot even
support a classical memory. Such a view is consistent with
Pirogov-Sinai~\cite{pirsin} type results where one shows that
well-separated ground states of infinite classical systems extend to
separated phases at sufficiently low temperatures.

A possible objection to the  previous reasoning, which was presented
in~\cite{AlickiH2006-drive}, is that the
mathematical framework of infinite systems, used to describe the 
thermodynamic limit, is too simplistic. 
One cannot exclude that observables which distinguish
between the degenerated ground states of the finite system but
disappear in the thermodynamic limit are long-lived in the presence of
a generic heat bath at sufficiently low temperatures. This would imply
that the expected equivalence of the infinite system equilibrium
picture and the dynamical picture
is not universal and that a new type of ``purely dynamical phase
transitions'' exists. In order to prove or disprove this claim one
should study finite supposedly self-correcting systems weakly coupled
to generic models of heat baths and search for possibly slowly relaxing
degrees of freedom. These could then support states of encoded qubits
for times exponentially increasing with $N$.

It is interesting to compare the problem of stable quantum memories to
another long-standing problem of quantum mechanics: the absence of
certain superpositions of quantum states for sufficiently large
systems. Well-known examples are Schr\"odinger cat states, the
appearance of a classical world, the problem of molecular structure,
chiral molecules, deformed nuclei, \dots~\cite{joo}\ It seems to be generally
accepted that the main ingredient for the solution of this problem is
the stability of quantum states of the observed system with respect to
the coupling to an environment, sometimes identified with a measuring
apparatus or even a conscious observer. As in the case of quantum
memories two approaches have been developed, a static and a dynamical
one. The first relies on dressing the states of the bare system. This
can produce disjoint stable states of the dressed system when using
environments with an infinite number of degrees of freedom and leads to
unobservable superpositions, at least for coupling constant larger then
a certain critical value. The mathematical mechanism is completely
similar to that which is responsible for the appearance of disjoint
KMS~states in the thermodynamic limit. The dynamical approach, on the
other hand, refers to decoherence phenomena which transform
superpositions of states into mixtures with respect to a certain
particularly stable pointer basis. The relations between both pictures
and their ranges of applicability are still a topic of debate.
Unfortunately, except for the spin-boson model~\cite{spo} no rigorous
result has been obtained.

The purpose of this paper is to start a systematic and rigorous
comparison of the infinite system equilibrium picture and the finite
system dynamical picture for models designed to support quantum
memories. We restrict ourselves to the 1D (quantum) Ising model and the
2D Kitaev model. Although there exist heuristic arguments that, in
spite of earlier claims concerning the 2D case~\cite{KitaevFT,
LloydRA1999-robcomp,LloydZ2002-top}, both
models cannot provide stable quantum memories~\cite{AlickiH2006-drive} it is
instructive to develop the basic mathematical tools for analysing these
relatively simple cases, hoping that they will be useful for 3D or 4D
models. Only the last case is presently considered by some experts to
be a proper candidate for a quantum memory~\cite{DennisKLP2002-4DKitaev}. One should
mention that formally the 1D quantum Ising model can be considered to
be the 1D Kitaev model and hence the parallel analysis of both could be
helpful for further studies.

The main simplifying feature of the Kitaev models is the absence of
wave propagation. This leads to a discrete spectrum of the Hamiltonian
with level spacings independent of the size of the system. This
property allows to apply the Markovian approximation in terms of Davies
weak coupling limit which reproduces fully the expected phenomenology
of an open system weakly coupled to a heat bath. The Davies semigroup
generators possess a simple local structure and behave well with
respect to the thermodynamic limit. Moreover, for Kitaev's model we can
determine the structure of the algebras which can encode qubits and we
obtain the dynamics of the qubit observables.

\section{Ground and equilibrium states of 1D Ising and 2D Kitaev models}

The physical substrate which is to carry the qubits has basic spin 
observables that are attached to the sites of a chain, a ring, a lattice
\dots\ We will describe their basic observables by Pauli matrices
$\sigma_j^\alpha$ where $j$ points at the site at which the spin lives and 
$\alpha = x,y,z$.  These basic observables obey the relations
\begin{align}
 &\sigma_j^x \sigma_j^y = i \sigma_j^z 
 \quad\text{and cyclic permutations of } x,y,z \\
 &\bigl(\sigma_j^\alpha\bigr)^\dagger = \sigma_j^\alpha 
  \qquad\text{and}\qquad 
  \bigl( \sigma_j^\alpha \bigr)^2 = \idty \\
 &[\sigma_j^\alpha, \sigma_k^\beta] = 0 \quad \text{for } j \ne k.
\end{align}

For systems with a finite number of spins the energy of the system is
encoded in the Hamiltonian and physically relevant states are ground
and thermal states. The first are determined by the eigenstates of the
Hamiltonian corresponding to the lowest energy and the second are given
by the Gibbs canonical density matrix at temperature $T$. The limit
$T\to0$ of the Gibbs state is a ground state. In degenerate cases this
limit is the average of all ground states and there are therefore
generally more ground states than just the low temperature limit of the
canonical Gibbs state.

Typically one is interested in quantities that have a nice scaling
behaviour with respect to $N$ and hence in the leading asymptotic
behaviour. This is precisely what the formalism of systems with an
infinite number of particles catches: it provides right away a
framework that allows to define and compute the asymptotic ground and
equilibrium states. Obviously the total energy of an infinite systems
is not a sensible observable and one characterises ground and
equilibrium states in terms of the interactions between the spins. In
fact sufficiently local interactions generate a well-behaved dynamics
on the infinite system given by a group of automorphisms $\{\alpha_t
\mid t\in\Rl\}$ of the observables. The generator of this group is a
derivation $\delta$ determined by $\alpha_t = \exp(it\delta)$. It
satisfies Leibniz's rule $\delta(X\,Y) = \delta(X)\,Y + X\,\delta(Y)$
and is self-adjoint $\delta(X^\dagger) = - \bigl( \delta(X) 
\bigr)^\dagger$. For a finite system with Hamiltonian $H$ one has 
$\delta(\cdot) = [H,\cdot]$. 

There are various ways of introducing ground and equilibrium states of
an infinite system, such as thermodynamical limits of local Gibbs
states possibly with boundary conditions or as states maximising the
entropy density for a given internal energy density. We shall use the
characterisation in terms of the dynamics. This leads to the ground
state equation and the KMS~condition. For finite systems these
conditions return the usual notions of ground state and canonical Gibbs
state at a given temperature.
  
A state $\omega$ satisfies the ground state equation if
\begin{equation}
 \omega\bigl( X^\dagger \delta(X) \bigr) \ge 0.
\label{gs}
\end{equation}
It satisfies the KMS~condition at inverse temperature $\beta = 1/k_{\r
B}T > 0$ if there exists
for any choice of observables $X$ and $Y$ a function
\begin{equation}
 F_{X,Y}:  \{z\in\Cx \mid 0 \le \Im\g m \le\beta\} \to \Cx
\end{equation} 
such that $F_{X,Y}$ is analytic in the open strip 
$\{z\in\Cx \mid 0 < \Im\g m < \beta\}$, is bounded in the closed strip,
extends continuously to the lines $\Im\g mz=0$ and $\Im\g m=\beta$ and
satisfies for $t\in\Rl$
\begin{equation}
 F_{X,Y}(t) = \omega(X\alpha_t(Y))
 \qquad\text{and}\qquad
 F_{X,Y}(t+i\beta) = \omega(\alpha_t(Y)X).
\end{equation} 
Loosely speaking, the KMS~condition says that
\begin{equation}
 \omega(X\alpha_{i\beta}(Y)) = \omega(YX)
\label{kms2}
\end{equation}
at least if the expression $\alpha_{i\beta}(Y)$ makes sense. It
therefore links equilibrium states with the Heisenberg dynamics at
imaginary times. For finite systems when the dynamics is given by
\begin{equation}
 \alpha_t(X) = \r e^{itH} X \r e^{-itH}
\end{equation}
it follows readily from~(\ref{kms2}) that only the canonical Gibbs
state at inverse temperature $\beta$ satisfies the KMS~condition.

If the system has a local symmetry in the sense that there exists a
local unitary $U$, i.e.\ a unitary living on a finite number of sites,
such that
\begin{equation}
 \alpha_t\bigl( U\,X\,U^\dagger \bigr) = U\,\alpha_t(X)\, U^\dagger
\end{equation}
and if $\omega$ is a KMS~state, then also the transformed state X
$\mapsto \omega(U\,X\,U^\dagger)$ is a KMS~state at the same inverse
temperature $\beta$ and is unitarily equivalent to the original one. A
general property of KMS~states, see\cite{RB97}, implies that $\omega$
is actually invariant under the symmetry
\begin{equation}
 \omega(X) = \omega\bigl( U\, X\, U^\dagger \bigr).
\label{e2}
\end{equation} 
There is no such result for ground states as determined by
equation~(\ref{gs}).

We shall consider two models: the Ising ferromagnet and a model
introduced by Kitaev. Both models have a finite version on a ring and a
torus respectively. We shall also consider the infinite versions, they
live on a chain and on a staggered square lattice. The particle
interactions are in both cases nearest-neighbour and both models,
either finite or infinite, allow for ground states in a very strong
sense in that the states minimise all local interactions. This can be
seen as a complete absence of frustration.

\subsection{1D Ising ferromagnet}

The interaction for an Ising ferromagnet is given in terms of bonds    
\begin{equation}
 Z_b := \sigma_j^z \sigma_{j+1}^z,
 \quad \text{with } 
 b = \{j,j+1\}.
\end{equation}
In this expression $j$ is either a site of a ring or a chain.

Let us first consider the finite system of $N$ spins on a ring, meaning
that we identify the sites 1 and $N+1$. Clearly the bond spins are not
independent as they satisfy the relation
\begin{equation}
 \prod_b Z_b = \idty.
\label{1}
\end{equation}
Any $(N\!-\!1)$ independent bond observables --- one can omit an
arbitrary one due to~(\ref{1})--- correspond to the $\sigma^z$
observables of $(N\!-\!1)$ qubits. The algebra $\c A_{\r z}$ generated
by the bonds is Abelian but is not generated by a maximal set of
commuting observables in the algebra of the $N$ spins. In fact its
commutant $\c A_{\r z}'$ within the spin algebra is the product of a
qubit algebra and $\c A_{\r z}$. This can be seen by the following
direct construction. Choose
\begin{equation}
 \s X = \sigma_1^x \sigma_2^x \cdots \sigma_N^x, \quad
 \s Y = \sigma_1^y \sigma_2^x \cdots \sigma_N^x, \quad \text{and }
 \s Z = \sigma_1^z.
\label{isingqubit}
\end{equation}
The matrices $\s X, \s Y$ and $\s Z$ satisfy the same relations
as the Pauli matrices and commute with the algebra of the bond
observables. It can be checked that any element commuting with $\c
A_{\r z}$ belongs to the algebra generated by the observables of the
encoded qubit~(\ref{isingqubit}) and $\c A_{\r z}$.
 
Let us  now turn to the infinite system which lives on a chain. The
local observables are finite linear combinations of finite products of
single site spin observables and the quasi-local observables are
obtained by completing this local algebra with respect to the natural
norm. The completion of the algebra of the bond observables $\c A_{\r
z}$ is again commutative but this time the commutant within the
quasi-local algebra is also Abelian, it is generated by e.g.\
$\sigma_1^z$ and $\c A_{\r z}$. The reason for this simplification is
to be found in the disappearance of the infinite product of $\sigma^x$
observables in the limit $N\to\infty$ for the $\s X$ and $\s Y$
observables in~(\ref{isingqubit}). From a mathematical point of view
such a limit does not exist in the spin algebra, from a physical point
of view the distribution of such observables has no stable limit in any
physically reasonable state.

We shall examine the ground and equilibrium states of the Ising
Hamiltonian with local Hamiltonian
\begin{equation}
 H^{\r{Ising}}_\Lambda = - \sum_{b\subset\Lambda} Z_b 
\end{equation}
Here $\Lambda$ is any finite subset of the chain.

\subsubsection{Ground states}

To compute the ground states we can use the following 
property of a state: suppose that $X=X^\dagger$ such that
$X \le \idty$ and that $\omega(X) = 1$, then 
$\omega(YX) = \omega(XY) = \omega(Y)$ for any $Y$. Indeed, $0 \le \idty
- X$ and so we can extract the square root of $\idty - X$ and 
apply the Cauchy-Schwartz inequality
\begin{equation}
 \abs{\omega(Y(\idty - X))}^2 \le \omega(Y(\idty - X)Y^\dagger)\, 
 \omega(\idty - X) = 0.
\label{cs} 
\end{equation}
Similarly $\omega(XY) = \omega(Y)$. 

Suppose now that $\omega$ is a state on the Abelian algebra $\c A_{\r
z}$ generated by the bond observables $Z_b$ such that $\omega(Z_b) = 1$
and let $\tilde\omega$ be an extension of this state to the spin
algebra. As $Z_b \le \idty$, we can apply the argument of above and
conclude that for any element $X$ in the spin algebra and any bond $b$
$\tilde\omega(X\,Z_b) = \tilde\omega(Z_b\,X) = \tilde\omega(X)$. 

This implies that $\tilde\omega$ is a ground state in the sense
of~(\ref{gs}). It suffices to show that $\tilde\omega(X^*\delta(X)) \ge
0$ for any local observable $X$. For such observables the action of the
derivation $\delta$ is given by a finite sum of commutators
\begin{equation}
 \delta(X) = - \sum_{b\in\Lambda} [Z_b,X].
\end{equation}
Here $\Lambda$ is the set of bonds which have a non-empty intersection
with the dependency set of $X$. We now compute
\begin{align}
 \tilde\omega(X^\dagger \delta(X))
 &=\sum_{b\in\Lambda} \tilde\omega\bigl( X^\dagger X\, Z_b - X^\dagger
 Z_b\, X\bigr) \\
 &=\sum_{b\in\Lambda} \Bigl( \tilde\omega\bigl( X^\dagger X\bigr) -
 \tilde\omega\bigl(X^\dagger\, Z_b\, X\bigr) \Bigr) \\
 &=\sum_{b\in\Lambda} \tilde\omega\bigl( X^\dagger(\idty - Z_b)X\bigr)
 \ge 0.
\end{align}

To compute the ground state expectation of an arbitrary spin observable
for the finite system we consider arbitrary monomials in Pauli
matrices. Using 
\begin{equation}
 \tilde\omega(\sigma_j^x) = - \tilde\omega(Z_b\, \sigma_j^x\,
 Z_b) = -\tilde\omega(\sigma_j^x)
\label{2}
\end{equation}
for a site $j$ which belongs to the bond $b$ and the obvious extensions 
to arbitrary polynomials, we conclude that
only monomials belonging to the commutant of $\c A_{\r z}$ can have
non-zero expectation values. Moreover with $\s X$ as in~(\ref{isingqubit}) 
\begin{equation}
 \tilde\omega(\s X B) = \tilde\omega(\s X)\, \omega_0(B),\quad B\in\c A_{\r
 z},
\end{equation}
and similar expressions replacing $\s X$ by $\s Y$ or $\s Z$. The
state $\omega_0$ is the unique state on $\c A_{\r z}$ that assigns the
value 1 to all products of bond observables. We have therefore
determined all ground states of the finite model: they are products 
of an arbitrary state on the
qubit algebra and the state $\omega_0$.

The infinite system can be handled in exactly the same way, we now
obtain product states of a classical bit with the state $\omega_0$. In
particular, this implies that the set of ground states of the Ising
ferromagnet on an infinite chain is a simplex, consisting only
of mixtures of two extremal states. In principle, this could be used
to encode a classical bit. In fact, this is a bit too optimistic as it
is well-known for this model that at finite temperatures there is a
unique equilibrium state.

\subsubsection{Equilibrium states}

Determining the equilibrium states of the finite and infinite Ising
models as solutions of the KMS~condition can be carried out along the
same lines as in the previous section. Using for any site $j$ belonging
to a bond $b$
\begin{equation}
 Z_b\, \sigma_j^x\, Z_b = - \sigma_j^x
 \qquad\text{and}\qquad
 Z_b\, \sigma_j^y\, Z_b = - \sigma_j^y
\end{equation} 
and the general result about KMS~states mentioned in~(\ref{e2}) we
conclude that the only monomials that have non-zero expectation values
belong to the commutant of $\c A_{\r z}$. The solution of the
KMS~condition is the extension of the equilibrium state for the bond
observables to $\c A_{\r z}'$ assigning zero values to any observable
of the form $\s X B$, $\s Y B$ or $\s Z B$, $B\in\c A_{\r z}$ for
the finite system and zero value to $\sigma^z_1B$ for the infinite
system.

\subsection{Kitaev's model in 2D}

The finite version of Kitaev's model lives on a $K \times K$ lattice
with periodic boundary conditions (toroidal lattice), see the square
dots in Fig.~1. The spins live on
the edges of the lattice as shown by the black and white dots in
the figure and so we have $N = 2 K^2$ qubits. 
The interactions between the spins are given by
\emph{star} and \emph{plaquette} terms $X_s$ and  $Z_p$. A
star is a diamond whose vertical vertices lie on the lattice of black
dots, while the vertical vertices of a plaquette lie on the lattice of
white dots. The actual observables are then
\begin{equation}
 X_s = \prod_{j\in s} \sigma_j^x 
 \qquad\text{and}\qquad 
 Z_p = \prod_{j\in p} \sigma_j^z.
\end{equation}     
The names star and plaquette correspond to the squares and stars marked 
by thick lines in the figure.

\begin{center}
\unitlength .4mm 
\linethickness{0.4pt}
\ifx\plotpoint\undefined\newsavebox{\plotpoint}\fi 
\begin{picture}(118,139.75)(0,0)
\put(10,10){\circle*{4}}
\put(50,10){\circle*{4}}
\put(90,10){\circle*{4}}
\put(30,30){\circle{4}}
\put(70,30){\circle{4}}
\put(110,30){\circle{4}}
\put(10,50){\circle*{4}}
\put(50,50){\circle*{4}}
\put(90,50){\circle*{4}}
\put(30,70){\circle{4}}
\put(70,70){\circle{4}}
\put(110,70){\circle{4}}
\put(10,90){\circle*{4}}
\put(50,90){\circle*{4}}
\put(90,90){\circle*{4}}
\put(30,110){\circle{4}}
\put(70,110){\circle{4}}
\put(110,110){\circle{4}}
\put(48,28){\rule{4\unitlength}{4\unitlength}}
\put(88,28){\rule{4\unitlength}{4\unitlength}}
\put(8,28){\rule{4\unitlength}{4\unitlength}}
\put(8,68){\rule{4\unitlength}{4\unitlength}}
\put(48,68){\rule{4\unitlength}{4\unitlength}}
\put(88,68){\rule{4\unitlength}{4\unitlength}}
\put(8,108){\rule{4\unitlength}{4\unitlength}}
\put(48,108){\rule{4\unitlength}{4\unitlength}}
\put(88,108){\rule{4\unitlength}{4\unitlength}}
\thicklines
\put(32,30){\line(1,0){36}}
\thinlines
\dottedline(28,108)(12,92)
\dottedline(28,72.25)(28,72.25)
\dottedline(12,88)(28,72)
\dottedline(32,72)(48,88)
\dottedline(32,32)(48,48)
\dottedline(52,12)(68,28)
\dottedline(52,48)(68,32)
\dottedline(32,28)(48,12)
\thicklines
\put(10,90){\line(1,1){20}}
\put(30,110){\line(1,-1){20}}
\put(50,90){\line(-1,-1){20}}
\put(30,70){\line(-1,1){20}}
\put(10,90){\line(0,1){0}}
\put(50,50){\line(0,-1){40}}
\put(30,90){\makebox(0,0)[cc]{$p$}}
\put(53,34.25){\makebox(0,0)[cc]{$\ s$}}
\thinlines
\put(2,110){\line(1,0){26}}
\put(32,110){\line(1,0){36}}
\put(72,110){\line(1,0){36}}
\put(112,110){\line(1,0){6}}
\put(2,70){\line(1,0){26}}
\put(32,70){\line(1,0){36}}
\put(72,70){\line(1,0){36}}
\put(112,70){\line(1,0){6}}
\put(11.75,139.75){\line(-1,0){.25}}
\put(2,30){\line(1,0){26}}
\put(72,30){\line(1,0){36}}
\put(112,30){\line(1,0){6}}
\put(10,118){\line(0,-1){116}}
\put(50,118){\line(0,-1){68}}
\put(50,10){\line(0,-1){8}}
\put(90,118){\line(0,-1){116}}
\end{picture} \\
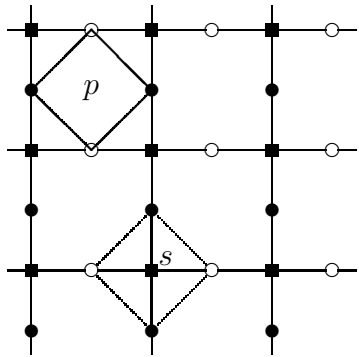
Fig 1: Kitaev's lattice
\end{center}

As stars and plaquettes have either 0 or 2 sites in common, $[X_s,
Z_p] = 0$. So, the algebras $\c A_{\r x}$ and $\c A_{\r z}$ generated by the
$X_s$ and $Z_p$ are Abelian. We shall denote by $\c A_{\r{xz}}$ the
algebra generated by $\c A_{\r x}$ and $\c A_{\r z}$. Because of the periodic
boundary conditions the star and plaquette observables are not
independent, they satisfy the relations
\begin{equation}
 \prod_s X_s = \idty 
 \qquad\text{and}\qquad 
 \prod_p Z_p = \idty, 
\end{equation}   
otherwise they are independent. As in the Ising case the algebra
$\c A_{\r{xz}}$
is not maximally Abelian within the spin algebra. Its commutant
consists of a product of two qubit algebras and $\c A_{\r{xz}}$. This can
again be seen quite explicitly by the following direct construction. 
Similarly to~(\ref{isingqubit}), one can introduce observables for two
encoded qubits
\begin{align}
 &\s Z_1 = \prod_{j\in c_1} \sigma_j^z,\qquad \s Z_2 = \prod_{j\in c_2} \sigma_j^z 
\notag\\ 
 &\s X_1 = \prod_{j\in c_1'} \sigma_{j'}^x, \qquad \s X_2 = \prod_{j\in c_2'} \sigma_{j'}^x.
\label{kitaevqubits}
\end{align}     
Here $c_1,\ c_1',\ c_2$ and $c_2'$ are the loops shown in
Fig.~2.

\begin{center}
\unitlength .4mm 
\linethickness{0.4pt}
\ifx\plotpoint\undefined\newsavebox{\plotpoint}\fi 
\begin{picture}(118,118)(0,0)
\put(10,10){\circle*{4}}
\put(50,10){\circle*{4}}
\put(90,10){\circle*{4}}
\put(30,30){\circle{4}}
\put(70,30){\circle{4}}
\put(110,30){\circle{4}}
\put(10,50){\circle*{4}}
\put(50,50){\circle*{4}}
\put(90,50){\circle*{4}}
\put(30,70){\circle{4}}
\put(70,70){\circle{4}}
\put(110,70){\circle{4}}
\put(10,90){\circle*{4}}
\put(50,90){\circle*{4}}
\put(90,90){\circle*{4}}
\put(30,110){\circle{4}}
\put(70,110){\circle{4}}
\put(110,110){\circle{4}}
\put(48,28){\rule{4\unitlength}{4\unitlength}}
\put(88,28){\rule{4\unitlength}{4\unitlength}}
\put(8,28){\rule{4\unitlength}{4\unitlength}}
\put(8,68){\rule{4\unitlength}{4\unitlength}}
\put(48,68){\rule{4\unitlength}{4\unitlength}}
\put(88,68){\rule{4\unitlength}{4\unitlength}}
\put(8,108){\rule{4\unitlength}{4\unitlength}}
\put(48,108){\rule{4\unitlength}{4\unitlength}}
\put(88,108){\rule{4\unitlength}{4\unitlength}}
\put(2,30){\line(1,0){26}}
\put(32,30){\line(1,0){36}}
\put(72,30){\line(1,0){36}}
\put(112,30){\line(1,0){6}}
\put(12,92.25){\line(0,1){0}}
\dottedline(28,112)(22,118)
\dottedline(32,112)(38,118)
\dottedline(12,92)(28,108)
\dottedline(48,92)(32,108)
\dottedline(12,88)(28,72)
\dottedline(48,88)(32,72)
\dottedline(12,52)(28,68)
\dottedline(48,52)(32,68)
\dottedline(12,48)(28,32)
\dottedline(32,32)(48,48)
\dottedline(12,12)(28,28)
\dottedline(48,12)(32,28)
\dottedline(12,8)(18,2)
\dottedline(48,8)(42,2)
\thicklines
\put(30,118){\line(0,-1){116}}
\put(2,30){\line(1,0){116}}
\put(96.75,28.5){\makebox(0,0)[ct]{$c_1'$}}
\put(31,10.75){\makebox(0,0)[lc]{$c_1$}}
\thinlines
\dottedline(10,118)(10,2)
\dottedline(90,118)(90,2)
\dottedline(2,110)(28,110)
\dottedline(32,110)(68,110)
\dottedline(72,110)(108,110)
\dottedline(112,110)(118,110)
\dottedline(2,70)(28,70)
\dottedline(32,70)(68,70)
\dottedline(72,70)(108,70)
\dottedline(112,70)(118,70)
\thicklines
\put(50,118){\line(0,-1){116}}
\put(2,50){\line(1,0){116}}
\put(53.25,15.5){\makebox(0,0)[cc]{$\ c_2$}}
\put(96.25,48.25){\makebox(0,0)[lt]{$\!\!c_2'$}}
\end{picture}
 \\
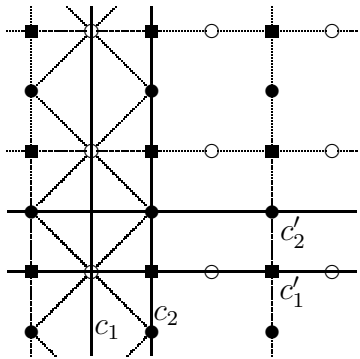
Fig~2: The loops in the toric lattice
\end{center}

Let us  now turn to the infinite system which lives on an infinite
staggered lattice. In contrast to the Ising case the commutant of the
algebra $\c A_{\r{xz}}$ in the spin algebra is now $\c A_{\r{xz}}$ itself. 
This can again be understood in terms of the non-existence of the
limits of the loop operators in~(\ref{kitaevqubits}) when $N\to\infty$. 
Indeed with increasing $N$, the length $K$ of the loops increases as
$N^{1/2}$.

The local Hamiltonians of the model are 
\begin{equation}
 H^{\r{Kitaev}}_\Lambda = - \sum_{\{s \mid s\subset\Lambda\}} X_s 
 - \sum_{\{p \mid p\subset\Lambda\}} Z_p.
\end{equation}
Here $\Lambda$ is any finite subset of the staggered lattice. 

\subsubsection{Ground states}

Let $\omega_0$ be the state on the Abelian algebra $\c A_{\r{xz}}$ generated by the $\c A_{\r x}$ and
$\c A_{\r z}$ such that
\begin{equation}
 \omega_0(X_s) = \omega_0(Z_p) = 1.
\label{3} 
\end{equation} 

Using the same arguments as for the Ising model we obtain that
$\omega_0$ extends to a product state of a two qubit algebra and $\c
A_{\r{xz}}$
for the finite model. For the infinite model there is only a unique
extension $\tilde\omega$ as we argue in the following lines and this
has, by unicity, to be pure. As
before, we compute the expectation values of a finite product of
elementary spin observables using~(\ref{cs}). Consider a product of
Pauli matrices living at the sites of a finite set $\Lambda$. Consider
first the point of $\Lambda$ farthest to the North and also, on the row
to which it belongs, farthest to the East. Suppose that this point
belongs to the lattice of black dots. An argument similar to that
in~(\ref{2}) kills the expectation of the observable unless it carries
a $\sigma^z$ Pauli matrix. Suppose it does and consider then the point
on the lattice of dots, lying South-East of the first. Unless this
carries also a $\sigma^z$ spin the expectation also vanishes because of
anticommutation with a suitably chosen plaquette observable. If both
points carry such $\sigma^z$ matrices, then they can be removed by
multiplying with a plaquette observable having the two points as
North-East boundary. A similar argument holds if our original point
belonged to the lattice of white dots. Continuing like this we see that the
only observables having non-zero expectation value are products of star
and plaquette observables, but all these have expectation 1 because
of~(\ref{cs}). So this model has a unique ground state and cannot be
used to encode even a single classical bit.

It remains to show that indeed a state satisfying~(\ref{3}) exists. It
suffices therefore to observe that we can map the star observables on
spins of an Ising model on $\Ir^2$ and the plaquettes on these of a
similar Ising model. Doing so, the interaction in the Kitaev model just
becomes that of a couple of free Ising spin systems on $\Ir^2$, which
clearly admits such a ground state.

\subsubsection{Equilibrium states}

Probably the most important feature of the plaquette and star observables is
that 
\begin{equation}
 [X_s,X_{s'}] = [X_s,Z_p] = [Z_p,Z_{p'}] = 0
\end{equation} 
for all choices of
$s$, $s'$, $p$ and $p'$. Moreover, 
\begin{equation}
 X_s = X_s^\dagger,\quad 
 Z_p = Z_p^\dagger,\quad \text{and } 
 X_s^2 = Z_p^2 = \idty. 
\end{equation}
Therefore, for the infinite system, the algebra generated by
the $X_s$ and $Z_p$ is isomorphic to the algebra of continuous function on a 
configuration space which is a countable product of copies of Ising spaces
$\{\uparrow,\downarrow\}$. We can, moreover, choose the isomorphism in such
a way that each of the $X_s$ and $Z_p$ is mapped onto an Ising variable.

Let $\Lambda \subset \Ir^2$ be a finite set. Adding successively boundary
layers, we obtain a collection of sets 
\begin{align}
 &\Lambda(0) := \Lambda \\
 &\Lambda(1) := \Bigl( \bigcup_{s\cap\Lambda(0) \ne \emptyset} s \Bigr) 
 \bigcup \Bigl( \bigcup_{p\cap\Lambda(0) \ne \emptyset} p \Bigr) \\
 &\Lambda(2) := \Bigl( \bigcup_{s\cap\Lambda(1) \ne \emptyset} s \Bigr) 
 \bigcup \Bigl( \bigcup_{p\cap\Lambda(1) \ne \emptyset} p \Bigr) \\
 &\cdots
\nonumber
\end{align}
In the Heisenberg picture, a local observable $X$ initially living in a
finite set $\Lambda$ evolves after time $t$ into
\begin{equation}
 \alpha_t(X) := X + it\delta(X) + \frac{(it)^2}{2!}\delta^2(X) + \cdots
\end{equation}     
where the derivation $\delta$ is obtained from the local Hamiltonians
\begin{align}
 \delta(X) 
 &= \lim_{M\to\Ir^2} \bigl[ H_M^{\r{Kitaev}}, X\bigr] \\
 &= - \sum_{s\subset\Lambda(1)} [X_s, X] - \sum_{P\subset\Lambda(1)} [Z_p,X]
 \\
 &= \bigl[ H_{\Lambda(1)}^{\r{Kitaev}}, X \bigr].
\end{align}
Using the commutation relations between the $X_s$ and $Z_p$, we see that
\begin{align}
 \delta^2(X) 
 &= - \sum_{s\subset\Lambda(2)} [X_s, \delta(X)] - \sum_{p\subset\Lambda(2)} 
 [Z_p,\delta(X)] \\
 &= - \sum_{s\subset\Lambda(2)} \sum_{s'\subset\Lambda(1)} [X_s,[X_{s'},X]] 
 - \cdots \\
 &= - \sum_{s'\subset\Lambda(1)} \sum_{s\subset\Lambda(2)} [X_{s'},[X_s,X]]
 - \cdots \\
 &= - \sum_{s'\subset\Lambda(1)} \sum_{s\subset\Lambda(1)} [X_{s'},[X_s,X]]
 - \cdots 
\end{align}
We therefore have for any $k=1,2,\ldots$
\begin{equation}
 \delta^k(X) = \bigl[ H_{\Lambda(1)}^{\r{Kitaev}}, \bigl[
 H_{\Lambda(1)}^{\r{Kitaev}}, \cdots \bigl[ H_{\Lambda(1)}^{\r{Kitaev}}, X \bigr]
 \cdots \bigr]\bigr]\quad \text{$k$ commutators}.
\end{equation}
Hence, for all $t\in\Rl$, $\alpha_t(X)$ lives in $\Lambda(1)$.  

The commutation of the $X_s$ and $Z_p$ also allows us to conclude that
for any $s$ and $p$ and any quasi-local observable $X$
\begin{equation}
 \alpha_t\Bigl( \bigl[ X_s,X \bigr] \Bigr) = \bigl[ X_s, \alpha_t(X) \bigr]
 \qquad\text{and}\qquad
 \alpha_t\Bigl( \bigl[ Z_p,X \bigr] \Bigr) = \bigl[ Z_p, \alpha_t(X) \bigr].
\label{e1}
\end{equation}

Because of~(\ref{e1}), given an $\alpha_t$-KMS~state $\omega$, 
for any choice of $s$ and $p$ and for
any $s\in\Rl$, also
\begin{equation}
 X \mapsto \omega\Bigl( \r e^{isX_s} X \r e^{-isX_s} \Bigr)
 \qquad\text{and}\qquad
 X \mapsto \omega\Bigl( \r e^{isZ_p} X \r e^{-isZ_p} \Bigr)
\end{equation} 
are $\alpha_t$-KMS~states at the same inverse temperature $\beta$. All such
states are unitarily equivalent to the original one. By~(\ref{e2}), $\omega$ is
invariant under all these local unitary transformations 
\begin{equation}
 \omega(X) = \omega\Bigl( \r e^{isX_s} X \r e^{-isX_s} \Bigr) = 
 \omega\Bigl( \r e^{isZ_p} X \r e^{-isZ_p} \Bigr).
\label{e3}
\end{equation} 

We now compute the $\alpha_t$-KMS~states. It suffices to compute the value
of $\omega$ on any monomial in the Pauli matrices, i.e.\ on any observable
of the kind
\begin{equation}
 X = \prod_{j\in\Lambda} \sigma^{\varepsilon_j}_j, \quad \varepsilon_j \in
 \{x,y,z\}.
\end{equation}
Using the invariances~(\ref{e3}) with the value $s=\pi$, we find that
\begin{equation}
 \omega(X) = \omega\Bigl( \prod_{j\in\Lambda} \sigma^{\varepsilon_j}_j \Bigr) 
 = 0, 
\end{equation}
unless $X$ is a product of star and plaquette observables. This implies that
$\omega$ is completely determined by its values on the commutative algebra
generated by the $\c A_{\r x}$ and $\c A_{\r z}$. The restriction of $\omega$ to this
algebra is, however, a classical equilibrium state of a free Ising
Hamiltonian at inverse temperature $\beta$. It is well-known that there is
only one such state which has a product structure fully determined by the
expectations
\begin{equation}
 \omega(X_s) = \omega(Z_p) = \tanh\bigl( \frac{\beta}{2} \bigr).
\end{equation}     

\section{The open Ising and Kitaev models}

We shall for both models consider noise originating from a coupling of
the atomic spins to a surrounding heat bath. In order to keep things as
simple as possible, we shall assume that each spin has its
private heat bath which is moreover independent from these of the
other spins. All baths are assumed to be identical. Moreover, we shall
restrict ourselves to the Markovian approximation applying directly the
weak coupling limit of Davies~\cite{Davies1974-WCL}.

Before treating the Ising and Kitaev models, we briefly sketch the
general setup and properties of Davies generators. A small system
is coupled to a collection of heat baths leading to the global Hamiltonian
\begin{equation}
 H = H^{\r{sys}} + H^{\r{bath}} + H^{\r{int}}
 \qquad\text{with}\qquad
 H^{\r{int}} = \sum_\alpha S_\alpha \otimes f_\alpha, 
\end{equation}
where the $S_\alpha$ are system operators and the $f_\alpha$ bath
operators. The main ingredients are the Fourier transforms $\hat
h_\alpha$ of the autocorrelation functions of the $f_\alpha$. The
function $\hat h_\alpha$ describes the rate at which the coupling is
able to transfer an energy difference $\omega$ from the bath to the
system. Often a minimal coupling to the bath is chosen, minimal in the
sense that the interaction part of the Hamiltonian is as simple as
possible but still addresses all energy levels of the system
Hamiltonian in order to produce finally an ergodic reduced dynamics.
The necessary and sufficient condition for ergodicity is~\cite{Spohn1977-ergod,Frigerio1978-ergod}
\begin{equation}
 \bigl\{ S_\alpha, H^{\r{sys}} \bigr\}' = \Cx\, \idty,
\end{equation}
i.e.\ no system operator apart from the multiples of the identity
commutes with all the $S_\alpha$ and $H^{\r{sys}}$. 

We begin by introducing the Fourier decompositions of the $S_\alpha$'s as they
evolve in time under the system evolution
\begin{equation}
 \r e^{itH^{\r{sys}}}\, S_\alpha\, \r e^{-itH^{\r{sys}}} 
 = \sum_\omega S_\alpha(\omega)\, \r e^{-i\omega t}. 
\end{equation}
Here the $\omega$ are the Bohr frequencies of the system Hamiltonian. From
self-adjointness we have the relation
\begin{equation}
 S_\alpha(-\omega) = S_\alpha(\omega)^\dagger.
\end{equation}
The weak coupling limit procedure then returns the following equation
for the evolution of the spin system in Heisenberg picture
\begin{align}
 \frac{dX}{dt} 
 &= i[H^{\r{sys}},X] + \c L_{\r{dis}}(X) =: \c L(X) 
\\
 \c L_{\r{dis}}(X)
 &= \frac{1}{2} \sum_\alpha \sum_\omega \hat h_\alpha(\omega) \Bigl(
 S_\alpha^\dagger(\omega)\, [X,S_\alpha(\omega)] + [S_\alpha^\dagger(\omega),X]\,
 S_\alpha(\omega) \Bigr)
\label{gen}
\end{align}
For thermal baths one has moreover the relation
\begin{equation}
 \hat h_\alpha(-\omega) = \r e^{-\beta\omega}\, \hat h_\alpha(\omega)
\end{equation} 
which is a consequence of the KMS~condition. The operator $\c L$
generates a semigroup of completely positive identity preserving
transformations of the spin system. It describes the reduced dynamics
in the Markovian approximation and enjoys the following properties
\begin{itemize}
\item
 The canonical Gibbs state with density matrix 
 \begin{equation}
  \rho_\beta = \frac{\r e^{-\beta H^{\r{sys}}}}{\tr \Bigl( \r e^{-\beta
  H^{\r{sys}}} \Bigr)}
 \end{equation}
 is a stationary state for the semigroup, i.e.\
 \begin{equation}
  \tr \Bigl( \rho_\beta\, \r e^{t\c L}(X) \Bigr) = \tr \bigl(
  \rho_\beta\, X \bigr).
 \end{equation}
\item
 The semigroup is relaxing, meaning that for any initial state $\rho$
 of the system
 \begin{equation}
  \lim_{t\to\infty}\ \tr \Bigl( \rho\, \r e^{t\c L}(X) \Bigr) = \tr \bigl(
  \rho_\beta\, X \bigr).
 \end{equation}
\item
 Furthermore, the generator satisfies the detailed balance condition,
 often called reversibility. Writing $\delta(X) :=
 [H^{\r{sys}},X]$,
\begin{equation}
 [\delta, \c L_{\r{dis}}] = 0
 \quad\text{and}\quad
 \tr \Bigl(\rho_\beta\, Y^\dagger\, \c L_{\r{dis}}(X) \Bigr) 
 = \tr \Bigl(\rho_\beta\, \bigl(\c L_{\r{dis}}(Y)\bigr)^\dagger\, X
 \Bigr).
\end{equation}
The last equation expresses the self-adjointness of the generator with
respect to the scalar product defined by the equilibrium state. The
space of observables equipped with the scalar product
\begin{equation}
 \<X,Y\>_\beta := \tr \rho_\beta\, X^\dagger\, Y 
\end{equation}
is called the Liouville space and the generator of the reduced dynamics
is a normal matrix on that space, i.e.\ the Hermitian and
skew-Hermitian parts of the generator commute.

\end{itemize} 

\subsection{Dissipative generators for Ising and Kitaev models}

The eigenstates and eigenvalues of the Ising Hamiltonian are
conveniently labelled by the excited bonds and one more dichotomic
variable which takes the twofold degeneracy into account. This last
could be the eigenvalues of $\s Z = \sigma_1^z$. An excited bond is often
called kink and can be seen as a quasi-particle. In the model on a
ring, kinks always appear in pairs due to the topology of the system.
The two dimensional Kitaev model will admit a very similar description
in terms of excited plaquette and star observables now called
anyons. 

More precisely, for the Ising model on a ring, the eigenstates of the
Hamiltonian are written as $|b_1, b_2, \ldots, b_{2n}; z\>$,
they have energy 
\begin{equation}
 H^{\r{Ising}}\, |b_1, b_2, \ldots, b_{2n}; z\> 
 = (2n - N)\, |b_1, b_2, \ldots, b_{2n}; z\>
\end{equation}
The interaction Hamiltonian is chosen as
\begin{equation}
 H^{\r{int}} = \sum_{j=1}^N \sigma^x_j \otimes f_j
\end{equation}
where $f_j$ is a self-adjoint field of the $j$-th bath and all the
baths are isomorphic.

The Bohr frequencies of the Ising Hamiltonian are $0,\ \pm2,\
\pm4,\ldots$. Only the Bohr frequencies $0$ and $\pm 2$ contribute to 
the Davies generator~(\ref{gen}) due to the choice of the
interaction. We compute the Fourier components of the $\sigma^x_j$'s
\begin{equation}
 \r e^{itH^{\r{spin}}}\, \sigma_j^x\, \r e^{-itH^{\r{spin}}} 
 = \r e^{-2it}\, a_j + \r e^{2it}\, a^\dagger_j + a_j^0.  
\end{equation}
The operators $a_j$ and $a_j^0$ are given in terms of projection
operators
\begin{equation}
 P_j^0 = \frac{1}{2}\, \bigl( \idty - Z_b Z_{b'} \bigr)
 \qquad\text{and}\qquad
 P_j^{\pm} = \frac{1}{4}\, \bigl( \idty \mp Z_b \bigr)\,
 \bigl( \idty \mp Z_{b'} \bigr)
\end{equation}
with $j=b\cap b'$:
\begin{align}
 &a_j^0 = P_j^0\, \sigma_j^x\, P_j^0 = P_j^0\, \sigma_j^x = \sigma_j^x\, P_j^0
 \\ 
 &a_j = P_j^-\, \sigma_j^x\, P_j^+ = P_j^-\, \sigma_j^x = \sigma_j^x\,
 P_j^+.
\end{align}
The Davies operators allow for an interpretation in terms of kinks.
Consider two bonds $b$ and $b'$ having a site $j$ in common. The
operator $a_j^0$ kills states for which both $b$ and $b'$ are
either empty or occupied and exchanges the occupations if only one of
them is. The operators $a_j^+$ and $a_j^-$ create and annihilate states
with two kinks meeting at $j$ respectively and kill the other states.
This leads us to the master equation in Heisenberg picture 
\begin{align}
 \frac{dX}{dt}
 &= i[H^{\r{Ising}},X] 
\nonumber\\ 
 &+ \frac{1}{2} \sum_{j=1}^N \Bigl\{ \hat h(2)\, \Bigl(
 a_j^\dagger\, [X,a_j] + [a_j^\dagger,X]\, a_j + \r e^{-2\beta}\, a_j\,
 [X,a_j^\dagger] + \r e^{-2\beta}\, [a_j,X]\, a_j^\dagger \Bigr) 
\nonumber\\ 
 &\phantom{+ \frac{1}{2} \sum_{j=1}^N \Bigl\{\ }- \hat h(0)\, 
 [a_j^0, [a_j^0,X]] \Bigr\}
\end{align}

For Kitaev's model on a torus, the eigenstates of the
Hamiltonian are written as 
$|p_1, p_2, \ldots, p_{2m}; s_1, s_2, \ldots, s_{2n};
z_1,z_2\>$. The labels $z_1$ and $z_2$ refer to the eigenvalues of the
$z$-components of the encoded qubits while the $p_j$ and $s_k$
point to the excited plaquette and star observables. Such an eigenstate
has energy 
\begin{align}
 &H^{\r{Kitaev}}\, |p_1, p_2, \ldots, p_{2m}; s_1, s_2, \ldots, s_{2n};
z_1,z_2\> 
\nonumber\\  
 &\qquad = (2m + 2n - N)\, |p_1, p_2, \ldots, p_{2m}; s_1, s_2, \ldots, s_{2n};
z_1,z_2\>
\end{align}
The interaction Hamiltonian is chosen as
\begin{equation}
 H^{\r{int}} = \sum_{j=1}^N \sigma^x_j \otimes f_j 
 + \sum_{j=N+1}^{2N} \sigma^z_j \otimes f_j
\end{equation}
where $f_j$ is a self-adjoint field of the $j$-th bath and all the
baths are isomorphic.

The Bohr frequencies of the Kitaev Hamiltonian are $0,\ \pm2,\
\pm4,\ldots$. Again, only the Bohr frequencies $0$ and $\pm 2$ contribute to 
the Davies generator~(\ref{gen}) due to the choice of the
interaction and we compute the Fourier decompositions of the spins as they
evolve in time
\begin{align}
 &\r e^{itH^{\r{spin}}}\, \sigma_j^x\, \r e^{-itH^{\r{spin}}} 
 = \r e^{-2it}\, a_j + \r e^{2it}\, a^\dagger_j + a_j^0 
\\  
 &\r e^{itH^{\r{spin}}}\, \sigma_j^z\, \r e^{-itH^{\r{spin}}} 
 = \r e^{-2it}\, b_j + \r e^{2it}\, b^\dagger_j + b_j^0.  
\end{align}
The operators $a$ and $b$ are again given in terms of projection
operators
\begin{equation}
 P_j^0 = \frac{1}{2}\, \bigl( \idty - Z_p Z_{p'} \bigr)
 \qquad\text{and}\qquad
 P_j^{\pm} = \frac{1}{4}\, \bigl( \idty \mp Z_p \bigr)\,
 \bigl( \idty \mp Z_{p'} \bigr)
\end{equation}
with $j=p\cap p'$. We need a second set of projections associated
with the stars
\begin{equation}
 R_j^0 = \frac{1}{2}\, \bigl( \idty - X_s X_{s'} \bigr)
 \qquad\text{and}\qquad
 R_j^{\pm} = \frac{1}{4}\, \bigl( \idty \mp X_s \bigr)\,
 \bigl( \idty \mp X_{s'} \bigr)
\end{equation}
with $j=s\cap s'$. Then
\begin{align}
 &a_j^0 = P_j^0\, \sigma_j^x\, P_j^0 = P_j^0\, \sigma_j^x = \sigma_j^x\, P_j^0
 \\ 
 &a_j = P_j^-\, \sigma_j^x\, P_j^+ = P_j^-\, \sigma_j^x = \sigma_j^x\,
 P_j^+.
\end{align}
and
\begin{align}
 &b_j^0 = R_j^0\, \sigma_j^z\, R_j^0 = R_j^0\, \sigma_j^z = \sigma_j^z\,
 R_j^0 \\ 
 &b_j = R_j^-\, \sigma_j^z\, R_j^+ = R_j^-\, \sigma_j^z = \sigma_j^z\,
 R_j^+.
\end{align}

The Davies operators allow for a similar interpretation as in the Ising
case in terms of anyons.
This leads us finally to the master equation in Heisenberg picture 
\begin{align}
 \frac{dX}{dt}
 &= i[H^{\r{Kitaev}},X] 
\nonumber\\ 
 &+ \frac{1}{2} \sum_{j=1}^N \Bigl\{ \hat h(2)\, \Bigl(
 a_j^\dagger\, [X,a_j] + [a_j^\dagger,X]\, a_j + \r e^{-2\beta}\, a_j\,
 [X,a_j^\dagger] + \r e^{-2\beta}\, [a_j,X]\, a_j^\dagger \Bigr) 
\nonumber\\ 
 &\phantom{+ \frac{1}{2} \sum_{j=1}^N \Bigl\{\ }- \hat h(0)\, 
 [a_j^0, [a_j^0,X]] \Bigr\} 
\nonumber\\
 &+ \frac{1}{2} \sum_{j=1}^N \Bigl\{ \hat h(2)\, \Bigl(
 b_j^\dagger\, [X,b_j] + [b_j^\dagger,X]\, b_j + \r e^{-2\beta}\, b_j\,
 [X,b_j^\dagger] + \r e^{-2\beta}\, [b_j,X]\, b_j^\dagger \Bigr) 
\nonumber\\ 
 &\phantom{+ \frac{1}{2} \sum_{j=1}^N \Bigl\{\ }- \hat h(0)\, 
 [b_j^0, [b_j^0,X]] \Bigr\}.
\end{align}

Both the Davies generators for the Ising and Kitaev models extend to
generators of completely positive identity preserving semigroups on the
infinite chain and the infinite staggered lattice. This easily follows from
the locality of the elements in the generator and the non-propagation
of both the Hamiltonian and dissipative parts in the generators.  

\subsection{Dynamics of the qubit observables}

In order to treat both the Ising and Kitaev models in a unified way it
is useful to introduce the notation $\c A_{\r{ab}}$ for the Abelian
algebras of observables. In the Ising case $\c A_{\r{ab}} = \c A_{\r z}$
and for the Kitaev model $\c A_{\r{ab}} = \c A_{\r{xz}}$. In both cases
we denote by $\c C$ the commutant of $\c A_{\r{ab}}$ in the respective
spin algebras, so $\c C = \c Q\, \c A_{\r z}$ for Ising and $\c C = \c
Q_1\, \c Q_2\, \c A_{\r xz}$ for Kitaev. The algebras $\c Q$ and $\c
Q_{1,2}$ are the qubit algebras introduced in~(\ref{isingqubit}) 
and~(\ref{kitaevqubits}).
So, e.g.\ in the Ising case any element of $\c C$ can be written as
linear combination of $\idty$ and the qubit observables $\s X$, 
$\s Y$ and $\s Z$ with coefficients in $\c A_{\r{ab}}$. Let us 
denote generically by $\{\s Q\}$ the bases of the
one or two-qubit spaces, so $\{\s Q\} =
\{\idty, \s X, \s Y, \s Z\}$ in the Ising case. 

We consider first more general qubit observables $\{\tilde{\s Q}\} =
\{\idty, \tilde{\s X}, \tilde{\s Y}, \tilde{\s Z}\}$ which enjoy
the following stability property: $[\tilde{\s Q},A] = 0$ for all
$A\in\c A_{\r{ab}}$, i.e.\ $\{\tilde{\s Q}\} \subset \c C$. The
physical meaning of this condition is that the $\{\tilde{\s Q}\}$ are
not only invariant under the Ising or Kitaev dynamics but also under a
whole class of perturbed dynamics of the form
\begin{equation}
 H^{'\r{Ising}} := - \sum_b J_b\, Z_b
 \qquad\text{and}\qquad
 H^{'\r{Kitaev}} := - \sum_s J_s^x \, X_s - \sum_p
 J_p^z \, Z_p.
\end{equation}      
In these formulas, the coupling constants are arbitrary but strictly
larger than 0. Remark that these perturbed models return precisely the
same ground states as the original ones. In fact it will suffice to
consider encodings of the type
\begin{equation}
 \tilde{\s Q} = \s Q \, F_{\s Q}, \quad F_{\s Q} \in \c A_{\r{ab}} 
\end{equation}
where $\{\s Q\}$ is the choice mentioned above. In order to satisfy the
qubit relations we have to impose on the functions $F$ that
\begin{equation}
 F_{\s X}\, F_{\s Y} = F_{\s Z},\quad
 F_{\s Q} = F_{\s Q}^\dagger,\quad
 F_{\s Q}^2 = \idty,\quad\text{and }
 F_{\idty} = \idty.
\end{equation}

In order to quantify the quality of a memory we are looking for the
shortest decay time of the time autocorrelation functions
\begin{equation}
 \<\tilde{\s Q}(t),\tilde{\s Q}\>_\beta
 = \tr \Bigl( \rho_\beta\, \r e^{t\c L}\bigl( \tilde{\s Q}\bigr)\,
\tilde{\s Q}\Bigr) 
 = \tr \Bigl( \rho_\beta\, \r e^{t\c L_{\r{dis}}}\bigl( \tilde{\s Q}\bigr)\,
\tilde{\s Q}\Bigr) 
 = \norm{\tilde{\s Q}(t/2)}^2_\beta.
\end{equation} 
This chain of equalities follows from the properties of the generator
of the reduced dynamics, derived at the beginning of Section~3. The
time autocorrelation functions measure how well a system encoded with a
message at time 0 has retained its information up to time $t$. The rightmost
$\tilde{\s Q}$ in the product can be seen as a polarisation of the system
corresponding to writing a message at $t=0$ while the leftmost is the
readout at time $t$.

We shall now compute the action of the reduced dynamics on elements of
the form $\tilde{\s Q} = \s Q\, F_{\s Q}$ with $F_{\s Q}$ as above and
show that
\begin{equation}
 \r e^{t\c L} \bigl( \tilde{\s Q} \bigr) 
 = \s Q\, \r e^{t\tilde{\c L}_{\s Q}} \bigl(F_{\s Q}\bigr) 
 = \r e^{t\tilde{\c L}_{\s Q}} \bigl(F_{\s Q}\bigr)\, \s Q 
\label{semi2}
\end{equation}  
with
\begin{equation}
 \r e^{t\tilde{\c L}_{\s Q}}: \c A_{\r{ab}} \to \c A_{\r{ab}}
\end{equation}
a contracting semigroup, hence
\begin{equation}
 \<\tilde{\s Q}(t),\tilde{\s Q}\>_\beta 
 = \bigl\langle \r e^{t\tilde{\c L}_{\s Q}} \bigl( F_{\s Q} \bigr), F_{\s Q}
 \bigr\rangle_\beta 
\end{equation}
In order to verify~(\ref{semi2}) we need some relations
\begin{align}
 &\sigma_j^\alpha\, F\, \sigma_j^\alpha \in \c A_{\r{ab}} \text{
 whenever } F\in\c A_{\r{ab}} \\
 &\sigma_j^\alpha\, \s Q\, \sigma_j^\alpha = g_{\s Q}^\alpha(j)\, \s Q
 \text{ with } g_{\s Q}^\alpha(j) \in \{\pm 1\}.
\end{align}
Consider a typical contribution to the generator
\begin{equation}
 \c K \bigl( X \bigr) := a_j\, X\, a_j^\dagger 
 - {\textstyle \frac{1}{2}} a_j a_j^\dagger\, X 
 - {\textstyle \frac{1}{2}} X\, a_j a_j^\dagger 
\end{equation}
with 
\begin{equation}
 a_j = P_j^-\, \sigma_j^x\, P_j^+,\quad P_j^\pm \in \c A_{\r{ab}}.
\end{equation}
For $F \in \c A_{\r{ab}}$ we obtain
\begin{align}
 \c K \bigl( \s Q\,F \bigr)
 &= P_j^-\, \sigma_j^x\, \s Q\, \sigma_j^x\, \sigma_j^x\, F\,
 \sigma_j^x\, P_j^- - \s Q\, P_j^-\, F \\
 &= \s Q \Bigl( P_j \bigl( g^x_{\s Q}(j)\, \sigma_j^x\, F\, \sigma_j^x
 - F\bigr)\Bigr)
 =: \s Q\, \tilde{\c K}_{\s Q}\bigl(F\bigr).
\end{align}
Clearly $\tilde{\c K}_{\s Q}\bigl(F\bigr) \in \c A_{\r{ab}}$. It now suffices to
add all contributions in order to show~(\ref{semi2}).
In this way we obtain the following expressions for
the reduced generators $\tilde{\c L}_{\s Q}$
\begin{align}
 \tilde{\c L}^{\r{Ising}}_{\s Q} \bigl(F\bigr)
 &= \sum_{j=1}^N D^x_j\, \Bigl( g^x_{\s Q}(j)\, \sigma^x_j\, F\,
 \sigma^x_j - F \Bigr) \\ 
 \tilde{\c L}^{\r{Kitaev}}_{\s Q} \bigl(F\bigr)
 &= \sum_{j=1}^N \Bigl\{ D^x_j\, \Bigl( g^x_{\s Q}(j)\, \sigma^x_j\, F\,
 \sigma^x_j - F \Bigr) 
\nonumber\\
 &\phantom{= \sum_{j=1}^N \Bigl\{\ } + D^z_j\, \Bigl( g^z_{\s Q}(j)\,
 \sigma^z_j\, F\, \sigma^z_j - F \Bigr) \Bigr\} \\
 D_j^x 
 &= \hat h(-2)\, P^-_j + \hat h(2)\, P_j^+ + \hat h(0)\, P^0_j \\
 D_j^z 
 &= \hat h(-2)\, R^-_j + \hat h(2)\, R_j^+ + \hat h(0)\, R^0_j. 
\end{align}

Finally, let us analyse the evolution of some special observables in
Kitaev's model. 
Consider observables $\tilde{\s X}_1$ and  $\tilde{\s Z}_1$
of the first qubit given by 
\begin{equation}
\tilde{\s X}_1 = \s X_1 F^x, \quad 
\tilde{\s Z}_1 = \s Z_1 F^z,
\end{equation}
with $F^x\in {\c A}_x$,  $F^z\in {\c A}_z$, and $\s X_1,\s Z_1$
as in~(\ref{kitaevqubits}).
It suffices to consider one of them, e.g.\ $\tilde{\s Z}_1$. 
It turns out that the following Hamiltonian and coupling with the environment
generate the same time evolution of this observable 
\begin{equation}
 H^{\r{Kitaev}}_z = - \sum_{\r p } Z_{\r p} 
 \qquad\text{and}\qquad
 H^{\r{int}}=\sum_{j=1}^N \sigma_j^x\otimes f_j. 
\label{kit-z} 
\end{equation}
More precisely, we have 
\begin{equation}
 \r e^{{\c L}^{\r{Kitaev}}t}\, \tilde{\s Z}_1 =
 \r e^{{\c L}^{\r{Kitaev}_z}t}\, \tilde{\s Z}_1
\label{L-kit-z} 
\end{equation}
where ${\c L}^{\r{Kitaev}_z}$ is solely determined by Davies operators of
$a$-type
\begin{equation}
 a_j^0 = \sigma_j^xP_j^0
 \qquad\text{and}\qquad
 a_j = \sigma_j^xP_j^+.
\end{equation}
Similar relations hold for the second observable. 
This can be interpreted as follows: with this choice of basic observables,  
bit and phase evolve separately according to the same evolution up to 
an exchange of $\sigma_x$ and $\sigma_z$. 
This property is then inherited by the reduced 
generators. Such a separation is not surprising. 
It simply means that such a choice of qubit observables
is compatible with the structure of the quantum code behind Kitaev's model.
Namely, the subspace of ground states constitutes a so-called CSS~code~\cite{Steane1996-proc-css,ShorC1996-css}.
The characteristic feature of this class of codes is 
that the recovery procedure, aiming at restoring  the initial state
of the qubit after the attack by the noise, is divided into two stages:
first correcting bit errors and then phase errors. 
The correction procedures are again the same, modulo 
$\sigma_z \leftrightarrow \sigma_x$ exchange. 

The aim of this paper was to analyse the static and dynamical structure
of Kitaev's model in 2D, using the analogy with the Ising ring or
chain. More involved models in 3 and 4D should be addressable in a
similar way. A further topic is determining the spectral properties of
the reduced semigroup generators, especially the behaviour of
relaxation times as a function of the system size. Such questions are
important in order to decide whether such systems are reasonable
candidates for stable quantum memories. This will be dealt with in a
forthcoming paper.

\noindent
\textbf{Acknowledgements:}
We would like to thank J.~Preskill, P.~Horodecki and R.~Horodecki 
for stimulating discussions.
This work is supported by the Polish-Flemish bilateral grant BIL~05/11 (MF), 
the  Polish
Ministry of Science and Information Technology - grant
PBZ-MIN-008/P03/2003 (RA and MH), EU Integrated Project Qubit
Applications QAP -IST directorate contract 015848 (RA)
and EC IP SCALA (MH).

\end{document}